\documentclass[a4paper]{article}  

\usepackage{color}
\usepackage{amsmath}
\usepackage{bbold}

\usepackage{mathtools}
\usepackage{amssymb}
\usepackage{graphicx}
\usepackage{gensymb}
\usepackage{pdfpages}
\usepackage{enumitem}
\usepackage{pifont}
\usepackage{MnSymbol}
\usepackage{hyperref}
\usepackage{epsfig} 

\title{Quantum-granularity effect in the formation of supermixed solitons in ring lattices}

\author{Andrea Richaud $^{1,*}$ and Vittorio Penna $^{1}$\\
{\footnotesize $^{1}$ \quad  Department of Applied Science and Technology and u.d.r. CNISM, Politecnico di Torino, I-10129 Torino, Italy}\\
 {\footnotesize $^{*}$ \quad Correspondence:   {\tt \footnotesize andrea.richaud@polito.it}
}}

\begin{document}
\maketitle

\abstract{We investigate a notable class of states peculiar to a bosonic binary mixture featuring repulsive intraspecies and attractive interspecies couplings. We evidence that, for small values of the hopping amplitudes, one can access particular regimes marked by the fact that the interwell boson transfer occurs in a jerky fashion. This property is shown to be responsible for the emergence of a staircase-like structure in the phase diagram of a mixture confined in a ring trimer and to strongly resemble the mechanism of the superfluid-Mott insulator transition. Under certain conditions, in fact, we show that it is possible to interpret the interspecies attraction as an effective chemical potential and the supermixed soliton as an effective particle reservoir. Our investigation is developed both within a fully quantum approach based on the analysis of several quantum indicators and by means of a simple analytical approximation scheme capable of capturing the essential features of this ultraquantum effect.}

\section{Introduction}
\label{sec:Introduction}
The possibility to simultaneously Bose-condense two different boson components (whether they are two different chemical elements \cite{Two_species_heteronuclear}, two different isotopes \cite{Grim_Rev}, or two different spin states \cite{Two_spin_states}) and to trap them in optical lattices \cite{Catani_deg} has opened the door to the investigation of the intriguing phenomenology exhibited by the resulting ultracold mixtures. The behaviour of the latter is ruled by the competition among tunnelling processes (resulting from the spatial fragmentation of the condensates into separated wells), \textit{intra-} and the \textit{inter-} species couplings. Such interplay among different contributions in the overall energy balance of the system results, among the rest, in a rich scenario of mixing-demixing quantum phase transitions \cite{sep1,sep2,sep3,Angom,Kuldeep_Fluctuation}, in the emergence of novel quantum phases \cite{Buonsante_PRL,qe1,Belemuk}, in the possibility of entangling \cite{ent,Bruno} the two bosonic species, and in that of triggering peculiar dynamical regimes \cite{Gallemi_1,Noi_NJP}.

In particular, mixing-demixing transitions have been thoroughly described, in the case of small-size lattices, for \textit{repulsive} \cite{Bruno,NoiEntropy,NoiSREP,Noi_Zenesini,NoiJPC} and \textit{attractive} \cite{NoiPRA3} interspecies couplings. These analyses have highlighted rather complex quantum phase diagrams where various phases, differing in the degree of mixing and localization of the two bosonic species, are recognizable. The latter properties have been shown to be quantifiable by means of suitable indicators originally devised in the context of \textit{classical} fluids \cite{Camesasca}, but which can be easily and effectively extended to the case of \textit{quantum} gases. Mixing-demixing and mixing-supermixing transitions in ultracold bosonic mixtures, which involve the localization of the condensed species in different sites of the lattice, have also been shown to be strongly associated to the presence of criticalities in a number of quantum indicators. The latter include, but are not limited to, the functional dependence 
of the ground state energy on model parameters, the energy fingerprint (constituted by the structure of the first excited energy levels), and the degree of entanglement between the bosonic species of the mixture\cite{PennaLinguaJPB,NoiSREP,Noi_Zenesini,NoiPRA3,Bruno}. 

In this work, we shine light on a particular aspect of the phenomenology exhibited by two-species mixtures confined in optical lattices: the emergence of a quantum-granularity effect resulting from the combination of strong interspecies attractions and weak hopping amplitudes. In these circumstances, in fact, the minority species occupies just one of the available sites and tends to summon the majority species in the same site where it localizes (hence the term ``supermixed soliton"). Nevertheless, some bosons of the majority species do not enter the macroscopically occupied lattice site but remain spread in the remaining ones. The resulting ground-state configuration can be therefore regarded as the union of two parts: the supermixed soliton, which plays the role of a particle reservoir for the majority component, and the remaining sites, which constitute an effective single-species system featuring a variable number of particles. In this perspective, the interspecies attraction plays the role of an effective chemical potential, as it can finely control the number of bosons which are injected from (to) the supermixed soliton into (from) the remaining lattice sites.  Within this analogy, the jerky interwell transfer of majority bosons occurring in the system is discussed to strongly resemble the well-known mechanism underlying the superfluid-Mott insulator transition \cite{mix10,Jaksch,Greiner2002,Caleffi_Capone_2019}. Recently, there has been considerable interest toward the physics of few-body ultracold systems \cite{Sowinski2019one} since they allow to better understand fundamental properties of quantum systems. In the same spirit, a mesoscopic number of particles (instead of a macroscopic one) is employed throughout our analysis to better emphasize the emergence of the quantum granularity. Moreover, recent experimental advances \cite{Amico_Cataliotti,Amico_qubit} have demonstrated the possibility to realize systems of interacting atoms trapped in ring-shaped optical lattices, an achievement that opened the doors to the observation of important phenomena in 1D physics. 

%%%%%%%%%%%%%%%%%%%%%%%%%%%%%%%%%%%%%%%%%%
\section{The two-species model}
\label{sec:The_model}
A bosonic binary mixture trapped in a three-well potential (trimer) can be effectively described in terms of the Bose-Hubbard (BH) model. The relevant Hamiltonian,

$$
H= - T_a \sum_{j=1}^{3} \left(A_{j+1}^\dagger A_j +A_j^\dagger A_{j+1} \right) + \frac{U_a}{2} \sum_{j=1}^{3} N_j(N_j-1)-
$$
\begin{equation}
\label{eq:Hamiltoniana_BH_trimero}
 - T_b \sum_{j=1}^{3} \left(B_{j+1}^\dagger B_j +B_j^\dagger B_{j+1} \right)
  +\frac{U_b}{2} \sum_{j=1}^{3} M_j(M_j-1)  +W \sum_{j=1}^{3} N_j\, M_j, 
\end{equation}
in fact, can capture the ultra-quantum effects originating from the interplay between the spatial fragmentation of the two condensates and the competition among tunnelling ($T_a$ and $T_b$), \textit{intra-} ($U_a$ and $U_b$) and \textit{inter-} ($W$) species couplings \cite{NoiPRA2,NoiEntropy,NoiSREP,Noi_Zenesini,NoiJPC,NoiPRA3}. Operator $A_j$ ($A_j^\dagger$) destroys (creates) a species-a boson in the $j$-th site. The same holds for operators $B_j$ and $B_j^\dagger$ which, respectively, destroy and create a species-b boson in the $j$-th site. These operators satisfy standard bosonic commutators: $[A_j,\,A_k]=0$, $[A_j,\,A_k^\dagger]=\delta_{j,k}$, $[B_j,\,B_k]=0$, $[B_j,\,B_k^\dagger]=\delta_{j,k}$,  $[A_j,\,B_k]=[A_j^\dagger,\,B_k]=0$.  

Number operators $N_j:=A_j^\dagger A_j$ and $M_j=B_j^\dagger B_j$ respectively count the number of species-a and species-b bosons in the $j$-th site. Their sums, 

$$ 
    \sum_{j=1}^3 N_j =N_a, \qquad \qquad  \sum_{j=1}^3 M_j =N_b,
$$
represent two independent conserved quantities, meaning that $[H,\,N_a]=[H,\,N_b]=0$. The system we are going to investigate features a \textit{ring} geometry, and, for this reason, it is understood that $j=4\equiv 1$ in the summations of Hamiltonian (\ref{eq:Hamiltoniana_BH_trimero}). Moreover, in the following, we shall focus on those regimes featuring \textit{repulsive} intraspecies and \textit{attractive} interspecies couplings, that means $U_a>0$, $U_b>0$ and $W<0$.
%
%%%%%%%%%%%%%%%%%%%%%%%%%%%%%%%%%%%%%%%%%%
\section{A continuous-variable picture to investigate the formation of supermixed solitons}
\label{sec:Review_PRA3}
\subsection{The system phase diagram}
\label{sub:system_phase_diagram}
\begin{figure}[h!]
\centering
\includegraphics[width=0.8\linewidth]{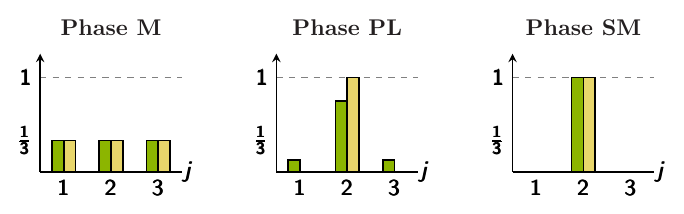}
\caption{Pictorial representation of  some states belonging to phases M, PL, and SM, respectively. Labels $1,\,2,\,3$ correspond to site numbers, while the vertical axis corresponds to (normalized) boson populations $x_{*,j}$ and $y_{*,j}$ characterizing the ground-state configuration. The majority (minority) species is depicted in green (yellow) and corresponds to the left (right) columns of the histograms in each panel. In phase M, the two bosonic species are mixed and uniformly distributed in the ring trimer; in phase PL, the minority species is highly localized, while the majority species occupies all the sites (although in a non-uniform manner); phase SM is characterized by supermixed solitons.}
\label{fig:Istogrammi}
\end{figure}

Bosonic binary mixtures trapped in ring lattices share, irrespective of the number of lattice sites, a rather general mechanism according to which, upon increasing the interspecies attraction $|W|$, the ground-state configuration undergoes deep changes \cite{NoiPRA3}. Basically, the two species are mixed and uniformly distributed in the lattice sites [mixed (M) phase] when $|W|$ is small enough. Conversely, when the latter becomes sufficiently negative, the minority species localizes in one site, while the majority species still occupies all sites, although in a non-uniform way [partially localized (PL) phase]. Eventually, further increasing $|W|$, both species localize in the same site, thus giving place to a state which goes under the name of ``supermixed soliton" [supermixed (SM) phase]. This scenario is pictorially illustrated in Figure \ref{fig:Istogrammi}.

\begin{figure}[h!]
\centering
\includegraphics[width=0.5\linewidth]{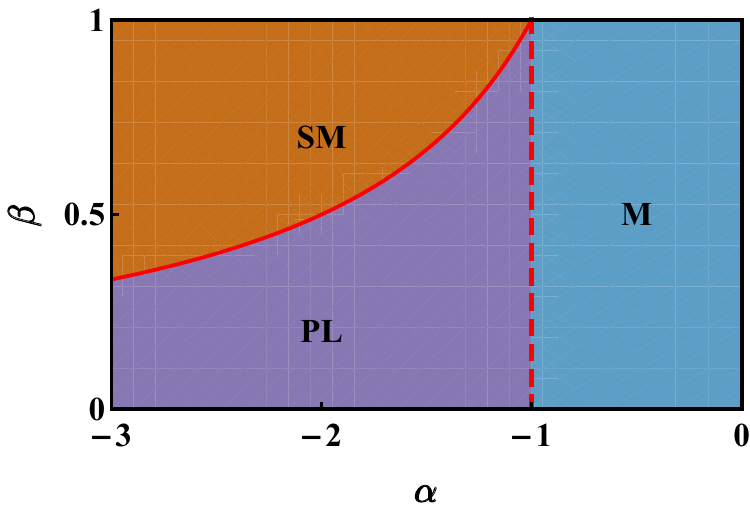}
\caption{Phase diagram of a (possibly asymmetric) two-species bosonic mixture confined in a $3$-well potential and featuring \textit{repulsive} intraspecies and \textit{attractive} interspecies interactions. Each phase is characterized by a specific functional dependence of the energy minimum (\ref{eq:V_*}) on effective model parameters (\ref{eq:alfa_beta}). Along the red dashed ($\alpha=-1$) and the red solid ($\beta=-1/\alpha$) lines, $V_*$ is not analytic, a circumstance which strongly suggests the occurrence of phase transitions. In the former (latter) case, it is the first (second) derivative of $V_*$ with respect to control parameter $\alpha$ to be discontinuous.}
\label{fig:Diagramma_di_fase}
\end{figure}

The analytic treatment developed in Ref. \cite{NoiPRA3} was based on the Continuous Variable (CV) Picture \cite{Spekkens,Javanainen,Ciobanu,Zin,Penna_Burioni}, a rather versatile approximation scheme which, under the assumption that the number of particles loaded in the system, $N_a$ and $N_b$, is large enough, allows one to turn the search for the ground state of Hamiltonian (\ref{eq:Hamiltoniana_BH_trimero}) into that for the global minimum of effective potential

\begin{equation}
\label{eq:V}
  V = \frac{1}{2} \sum_{j=1}^3 x_j^2 + \frac{\beta^2}{2} \sum_{j=1}^3 y_j^2 + \alpha\beta \sum_{j=1}^3 x_j y_j,
\end{equation}
an expression where variables 

\begin{equation}
\label{eq:Normalized_populations}
    x_j:=\frac{N_j}{N_a}, \qquad y_j:=\frac{M_j}{N_b}
\end{equation}
represent \textit{normalized} boson populations, and where only two effective parameters,

\begin{equation}
    \label{eq:alfa_beta}
    \alpha= \frac{W}{\sqrt{U_aU_b}}, \qquad \beta=\frac{N_b}{N_a} \sqrt{\frac{U_b}{U_a}},
\end{equation}
come into play \cite{Noi_Zenesini,NoiPRA3}. It is to be noted, in this regard, that, in the limit of large boson populations, not only the inherently \textit{discrete} variables $N_j$ and $M_j$ can be replaced with their \textit{continuous} counterparts $x_j$ and $y_j$, but also the contribution of tunnelling terms in potential (\ref{eq:V}) can be neglected (recall that tunnelling energy scales as $N_c$, while intra- and interspecies coupling energies scale as $N_c^2$, where $c=a,\,b$). The limit $N_a \gg 1$, $N_b \gg 1$, where $N_b/N_a=\mathrm{const}$, can be regarded as a sort of thermodynamic limit if one resorts to the statistical-mechanical framework discussed in \cite{Dynamical_Bifurcation,Oelkers} (see also Refs. \cite{NoiSREP,Noi_Zenesini,NoiPRA3}) and allows one to detect the presence of different phases in the $(\alpha,\,\beta)$ plane. These phases correspond to different classes of ground states of Hamiltonian (\ref{eq:Hamiltoniana_BH_trimero}), differ in the degree of mixing and localization of the two bosonic species and, at their borders, the energy corresponding to the configuration $(\Vec{x},\,\Vec{y})$ which minimizes (\ref{eq:V}), regarded as a function of control parameters $\alpha$ and $\beta$, features non-analiticities. More specifically, if the configuration $(\vec{x}_*,\vec{y}_*)$ constitutes the global minimum of potential (\ref{eq:V}), the associated energy,

\begin{equation}
\label{eq:V_*}
V_*:=V(\vec{x}_*,\vec{y}_*):=\min_{(\vec{x},\vec{y})\in \mathcal{R}}V(\vec{x},\vec{y}),
\end{equation}
where 

$$
\mathcal{R}=\left\{(\vec{x}_j,\vec{y}_j)\,:\, 0\le x_j,\,y_j\le 1, \quad \sum_{j=1}^3 x_j=\sum_{j=1}^3 y_j= 1 \right \},
$$
features different functional dependences in different regions of the ($\alpha,\,\beta$) plane and thus exhibits non-analiticities along the borders thereof. This circumstance strongly resembles the hallmark of quantum phase transitions \cite{Sachdev}. Figure \ref{fig:Diagramma_di_fase} illustrates the system phase diagram in the thermodynamic limit (mentioned above), while Table \ref{tab:Fasi_trimero} summarizes the ground state configuration and the associated energy in each of the three phases. 

\begin{table}[h]
\caption{Summary of the typical minimum-energy configuration and of the associated energy in each phase.}
\centering
\begingroup
\renewcommand{\arraystretch}{1.5} % Default value: 1
\begin{tabular}{c|c|c}
\hline
\textbf{Phase} & \(\displaystyle (\vec{x}_*,\vec{y}_*) \) & \(\displaystyle V_* \) \\ \hline
\textbf{M} & \begin{tabular}[c]{@{}c@{}}\(\displaystyle x_{*,j}=1/3 \quad \forall j \)\\ \\ \(\displaystyle y_{*,j}=1/3 \quad \forall j \)\end{tabular} & \(\displaystyle V_{*}^\mathrm{M}=\frac{1}{6}(\beta^2+2\alpha\beta+1)\) \\ \hline
\textbf{PL} & \begin{tabular}[c]{@{}c@{}}\(\displaystyle x_{*,i}=[1-2\alpha\beta]/3 \)\\  \(\displaystyle x_{*,j}=[1+\alpha\beta]/3  \quad \forall j\neq i \)\\ \\ \(\displaystyle y_{*,i}=1, \,\,\,  y_{*,j}=0 \,\,\, \forall j\neq i \)\end{tabular} & \begin{tabular}[c]{@{}c@{}}\(\displaystyle V_{*}^\mathrm{PL}=\frac{1}{6} [1+2\alpha\beta\) \\ \(\displaystyle + \beta^2 (3-2\alpha^2)]\)\end{tabular} \\ \hline
\textbf{SM} & \begin{tabular}[c]{@{}c@{}}\(\displaystyle x_{*,i}=1\)\\  \(\displaystyle x_{*,j}=0 \quad \forall j\neq i \)\\ \\ \(\displaystyle y_{*,i}=1, \,\,\, y_{*,j}=0 \,\,\, \forall j\neq i\)\end{tabular} & \(\displaystyle V_{*}^\mathrm{SM}=\frac{1}{2}(\beta^2+2\alpha\beta+1)\) \\ \hline
\end{tabular}
\endgroup
\label{tab:Fasi_trimero}
\end{table}

To conclude this Section, we remark that the presented study encompasses a rather extended portion of the parameters' space. With reference to definitions (\ref{eq:alfa_beta}), in fact, we have verified that no additional phases emerge for $\alpha<-3$, while the case of $\alpha>0$ has been thoroughly investigated in Refs. \cite{NoiSREP,Noi_Zenesini}. As regards parameter $\beta$, the choice $\beta\in[0,1]$ comes with no loss of generality in that, if $\beta$ happens to be bigger than $1$, one can always swap species labels and hence come back to the aforementioned interval $\beta\in[0,1]$. Notice also that the asymmetric role of species labels in the definition of $\beta$ [see formulas (\ref{eq:alfa_beta})] implicitly defines a \textit{majority} species, $a$, and a \textit{minority} species, $b$. 

\subsection{Some quantum indicators to characterize the different phases}
\label{sub:Quantum_indicators}
In order to better characterize the three possible phases exhibited by the system, one can make use of the ``entropy of mixing" and of the ``entropy of location", two indicators that are commonly used in Physical Chemistry \cite{Camesasca,Brandani} to quantify the degree of mixing and localization of chemical compounds. In the case of \textit{normal} fluids, they are defined as

\begin{equation}
    \label{eq:S_mix}
    S_{mix}(\vec{x},\,\vec{y})=-\frac{1}{2} \sum_{j=1}^L\left( x_j \log \frac{x_j}{x_j+y_j} +y_j \log \frac{y_j}{x_j+y_j} \right) 
\end{equation}

\begin{equation}
    \label{eq:S_loc}
    S_{loc}(\vec{x},\,\vec{y})= -\sum_{j=1}^L \frac{x_j+y_j}{2} \log \frac{x_j+y_j}{2}.
\end{equation}
where $x_j$ and $y_j$ are the molar fractions of the two compounds in the $j$-th spatial domain and $L$ represents the number thereof (spatial domains result from the discretization of the available volume). As we are dealing with \textit{quantum} fluids, the system ground state will be, in general, a superposition of different Fock states $|\vec{N},\vec{M}\rangle$, each one associated to a certain $S_{mix}$ and $S_{loc}$ which can be, in turn, determined by means of formulas (\ref{eq:S_mix}-\ref{eq:S_loc}) through the mapping (\ref{eq:Normalized_populations}) (of course, in our case, $L=3$ due to the presence of \textit{three} sites, which already constitute the most natural way to discretize the system's spatial domain). In this perspective, the quantum version of indicators (\ref{eq:S_mix}-\ref{eq:S_loc}) reads 

\begin{equation}
\label{eq:S_mix_tilde}
   \tilde{S}_{mix}:=\sum_{\vec{N},\vec{M}}^Q |c(\vec{N},\vec{M})|^2 S_{mix}(\vec{N},\vec{M}),
\end{equation}
\begin{equation}
\label{eq:S_loc_tilde}
   \tilde{S}_{loc}:=\sum_{\vec{N},\vec{M}}^Q |c(\vec{N},\vec{M})|^2 S_{loc}(\vec{N},\vec{M}), 
\end{equation}
where $Q=\frac{(N_a+2)!}{N_a!2!}\frac{(N_b+2)!}{N_b!2!}$
is the dimension of the Hilbert space of states associated to Hamiltonian (\ref{eq:Hamiltoniana_BH_trimero}) and 

\begin{equation}
\label{eq:c_k}
    c(\vec{N},\vec{M})= \langle\vec{N},\vec{M}|\psi_0\rangle 
\end{equation}
is the projection of the ground state $|\psi_0 \rangle$ onto Fock state $|\vec{N},\,\vec{M}\rangle =|N_1,\,N_2,\,N_3,\,M_1,\,M_2,\,M_3\rangle$. 

Other quantum indicators that can be used to detect the presence of different phases \cite{Sachdev} in the $(\alpha,\,\beta)$ plane are the ground-state energy

\begin{equation}
\label{eq:Ground_state_energy}
   E_0=\langle \psi_0 |H|\psi_0\rangle
   \end{equation}
and the first excited levels

\begin{equation}
\label{eq:Livelli_eccitati}
   E_i=\langle \psi_i |H|\psi_i\rangle.
\end{equation}
which indeed constitute a sort of energy fingerprint for quantum phases.

Eventually, in order to evaluate the degree of quantum correlation between the two species, one can introduce the entanglement entropy (EE) relevant to a bipartition of the system space of states in terms of species-$a$ and species-$b$ bosons \cite{PennaLinguaJPB,NoiEntropy,NoiSREP}. More specifically, the entanglement between the two quantum fluids reads

\begin{equation}
        EE = - \mathrm{Tr}_{a} (\hat{\rho}_{a}\, \log_2 \hat{\rho}_{a} ), 
\label{eq:EE}
\end{equation}
a formula representing the Von Neumann entropy of the reduced density matrix $\hat{\rho}_{a}=\mathrm{Tr}_{b} \left(\hat{\rho}_0\right)$ obtained, in turn, by tracing out the degrees of freedom of species-$b$ bosons from the ground-state density matrix $\hat{\rho}_0=|\psi_0\rangle\langle\psi_0|$.

\begin{figure}[h]
\centering
\includegraphics[width=1\linewidth]{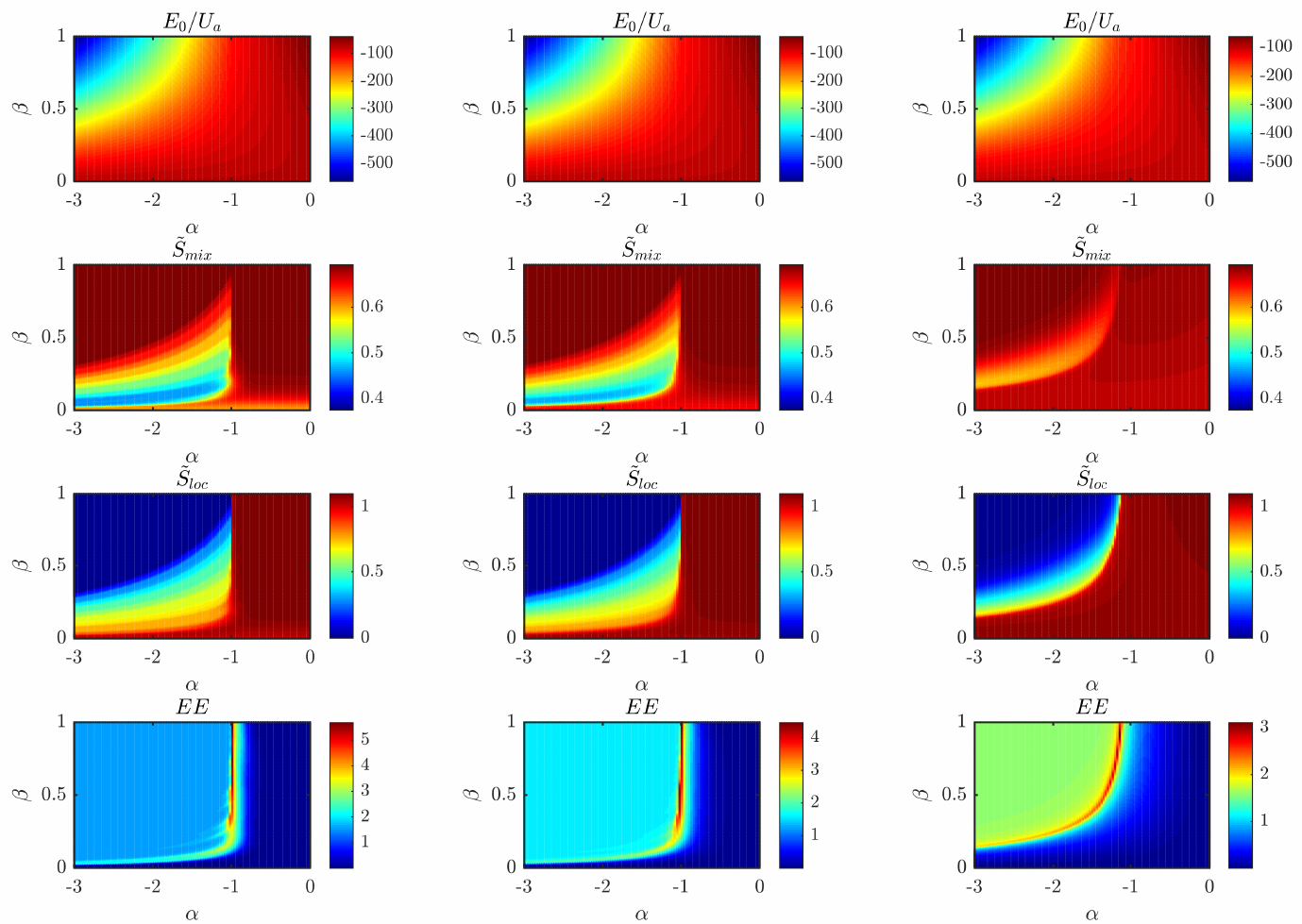}
\caption{Each row illustrates the behaviour of a genuinely quantum indicator as a function of model parameters $\alpha$ and $\beta$. Going from left to right, plots correspond to $T/U_a=0$, $0.02$ and $0.50$, where $T:=T_a=T_b$ . First row: ground-state energy $E_0/U_a$ (\ref{eq:Ground_state_energy}). Second row: quantum version of the entropy of mixing, $\tilde{S}_{mix}$ (\ref{eq:S_mix_tilde}). Third row: quantum version of the entropy of location $\tilde{S}_{loc}$ (\ref{eq:S_loc_tilde}). Fourth row: entanglement between the two condensed species, $EE$ (\ref{eq:EE}). Model parameters $N_a=N_b=15$, $U_a=1$, $U_b\in[0,1]\,\Rightarrow\, \beta\in[0,1]$ and $\alpha\in[-3,0]$ have been used. Each plot includes more than 20k points \cite{Hactar}, corresponding to as many numerical diagonalizations of Hamiltonian (\ref{eq:Hamiltoniana_BH_trimero}). }
\label{fig:Quantum_indicators_trimero_MI_SF}
\end{figure}

Figure \ref{fig:Quantum_indicators_trimero_MI_SF} illustrates the behaviour of these indicators, regarded as functions of effective model parameters (\ref{eq:alfa_beta}). It is possible to appreciate that the combined use of critical indicators $\tilde{S}_{mix}$ and $\tilde{S}_{loc}$ (see panels in the second and in the third row of Figure \ref{fig:Quantum_indicators_trimero_MI_SF}) allows one to clearly distinguish the different phases. It is worth noticing that, as the number of particles employed to perform the exact numerical diagonalization of Hamiltonian (\ref{eq:Hamiltoniana_BH_trimero}) is \textit{limited} ($N_a=N_b=15$), some finite-size effects are present, which affect the ``ideal" phase diagram illustrated in Figure \ref{fig:Diagramma_di_fase}. More prominently, the border between phase M and phase PL (line $\alpha=-1$ in the thermodynamic limit) has given way to a hyperbole-like border which allows phase M to invade the half-plane $\alpha<-1$ (of course, for sufficiently small values of $\beta$).

The last row of Figure \ref{fig:Quantum_indicators_trimero_MI_SF} illustrates the behaviour of $EE$. The transition M-PL can be easily recognized, while the border PL-SM cannot be appreciated (as already noticed in Ref. \cite{NoiPRA3}). 

Eventually, as it is visible in the first row of Figure \ref{fig:Quantum_indicators_trimero_MI_SF}, the ground-state energy $E_0$ as such does \textit{not} allow for a direct identification of the various phases because its non-analytic character is better highlighted by its first- and second-order derivatives. This aspect will be discussed in Section \ref{sec:Beyond} and illustrated in Figure \ref{fig:Quantum_indicators_trimero_derivatives_MI_SF}, where the \textit{derivatives} of $E_0$ with respect to $\alpha$ and $\beta$ (regarded, in turn, as functions thereof), are used to effectively reconstruct the phase diagram. Each column corresponds to a certain value of the tunnelling amplitudes $T_a$ and $T_b$. Going from left to right, the latter increase, a circumstance which favours boson delocalization and determines the blurring of the phase diagram illustrated in Figure \ref{fig:Diagramma_di_fase} (see Ref. \cite{NoiPRA3} for further details).

%%%%%%%%%%%%%%%%%%%%%%%%%%%%%%%%%%%%%%%%%%
\section{Beyond the continuous-variable picture: emergence of the quantum granularity}
\label{sec:Beyond}

The analytic treatment reviewed in Section \ref{sec:Review_PRA3} and based on the CV picture allows one to find all the phases that are possibly exhibited by the two-species mixture in a rather straightforward way. The resulting phase diagram (see Figure \ref{fig:Diagramma_di_fase}) and the associated characteristic quantities (see Table \ref{tab:Fasi_trimero}) provide a full overview of the different ways in which the two quantum fluids can rearrange among available sites and allows one to recognize critical lines in the $(\alpha,\,\beta)$ plane.

Nevertheless, this semiclassical approximation scheme cannot accurately describe the ultraquantum effects exhibited by the system when boson populations $N_a$ and $N_b$ are \textit{finite} and tunnelling processes very weak. In these cases, in fact, a small variation of control parameters $\alpha$ and $\beta$ [see formulas (\ref{eq:alfa_beta})] may \textit{not} result in a smooth variation of the system's ground state. 

To better clarify this circumstance, we begin with considering the central and the right panels of Figure \ref{fig:Istogrammi}. In the thermodynamic limit, one loses track of the quantum granularity characterizing bosonic particles, and phase PL can be thought of as a collection of states which, upon increasing $|\alpha|$, \textit{smoothly} approach the supermixed-soliton configuration. Therefore, in this scenario, the majority species \textit{gradually} localizes upon increasing the interspecies attraction, meaning that the outer green bars in the central panel of Figure \ref{fig:Istogrammi} are \textit{smoothly} reabsorbed by the emerging supermixed soliton. 

\begin{figure}[h!]
\centering
\includegraphics[width=0.6\linewidth]{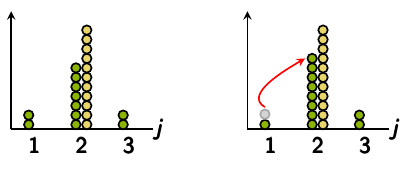}
\caption{Pictorial representation of the discrete character of the interwell boson exchange. Left panel: macroscopic configuration of the system for a certain choice of model parameters. A small variation of control parameters $\alpha$ and $\beta$ may (right panel) or may not modify it. The fact that the supermixed soliton can gain or loose a boson at a time upon varying a control parameter is what we mean with the term ``quantum granularity".}
\label{fig:Istogrammi_con_bolle}
\end{figure}

In this Section, both by means of exact numerical computations and by developing a suitable analytic framework, we show that this smooth and elementary picture is no longer valid for finite values of boson populations ($N_a$ and $N_b$) and for sufficiently low values of $T_a$ and $T_b$. In these regions of the parameters' space, the discrete character of the interwell boson-exchange mechanism emerge and the system discloses some new effects ensuing from the granularity of its constituents. Figure \ref{fig:Istogrammi_con_bolle} provides a pictorial representation of this phenomenology.     

\subsection{Exact numerical results}
\label{sub:Exact_results}
\begin{figure}[h]
\centering
\includegraphics[width=1\linewidth]{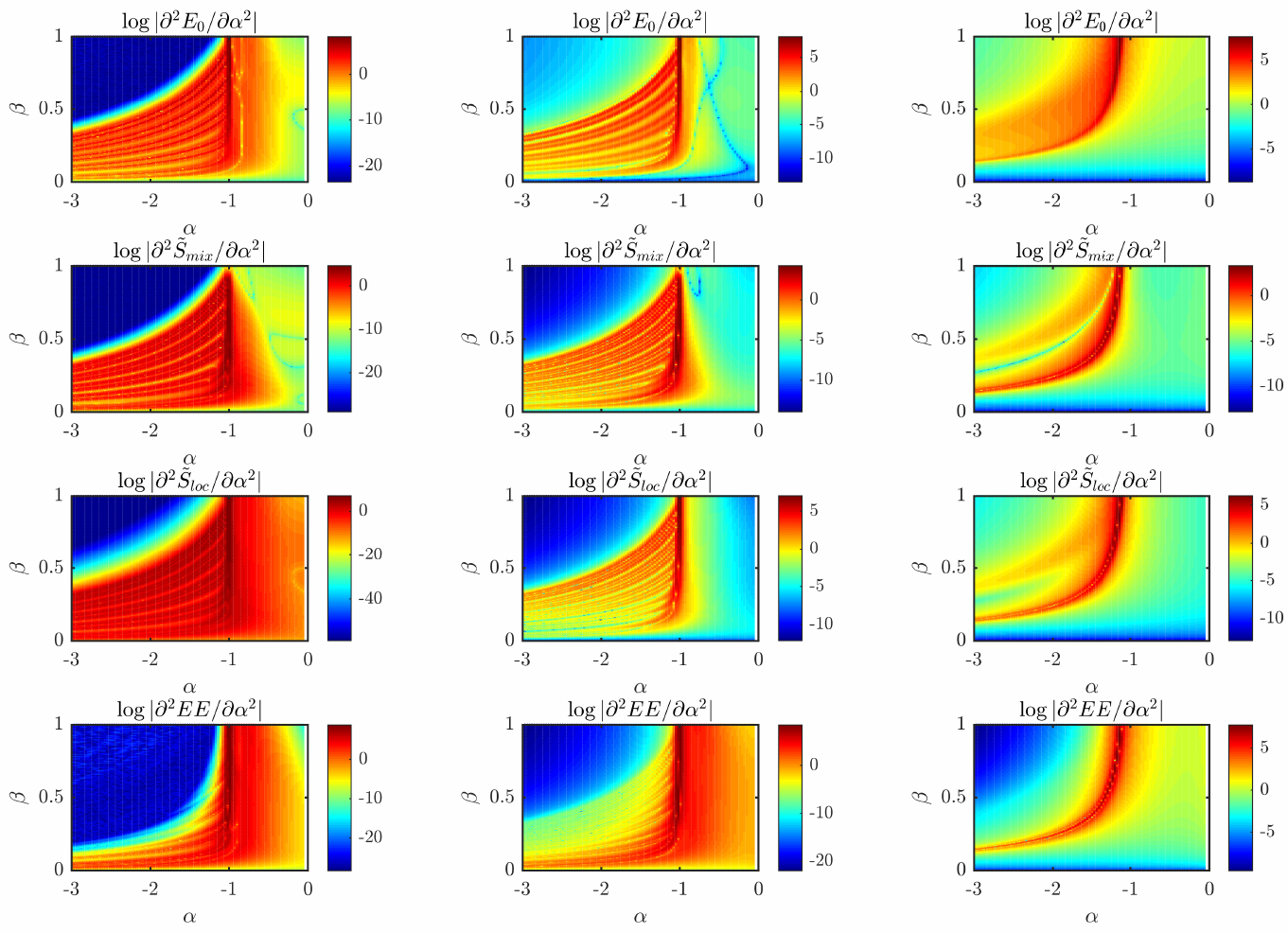}
\caption{Each row illustrates the behaviour of a genuinely quantum indicator as a function of model parameters $\alpha$ and $\beta$. Going from left to right, plots correspond to $T/U_a=0$, $0.02$ and $0.50$, where $T:=T_a=T_b$. First row: Second derivative of the ground-state energy $E_0$. Second row: second derivative of the quantum version of the entropy of mixing, $\tilde{S}_{mix}$. Third row: second derivative of the quantum version of the entropy of location $\tilde{S}_{loc}$. Fourth row: second derivative of the entanglement between the two condensed species, $EE$. Model parameters $N_a=N_b=15$, $U_a=1$, $U_b\in[0,1]\,\Rightarrow\, \beta\in[0,1]$ and $\alpha\in[-3,0]$ have been used. Each plot includes more than 20k points \cite{Hactar}, corresponding to as many numerical diagonalizations of Hamiltonian (\ref{eq:Hamiltoniana_BH_trimero}).}
\label{fig:Quantum_indicators_trimero_derivatives_MI_SF}
\end{figure}

The emergence of the aforementioned ``quantum granularity" in the phenomenology of the discussed system can be appreciated by resorting to the quantum indicators already introduced in Section \ref{sub:Quantum_indicators} and including the ground-state energy, the entropy of mixing, the entropy of location, and the entanglement entropy. In Figure \ref{fig:Quantum_indicators_trimero_derivatives_MI_SF}, we illustrate their second derivatives with respect to control parameter $\alpha$, where, for the sake of simplicity, we have set $T:=T_a=T_b$. It is clear that, in the region of the $(\alpha,\,\beta)$ plane corresponding to phase PL, a staircase-like structure is present for sufficiently low values of $T$ (see left and central columns of Figure \ref{fig:Quantum_indicators_trimero_derivatives_MI_SF}). Conversely, this peculiar property is \textit{absent} when tunnelling is large enough (see right column of Figure \ref{fig:Quantum_indicators_trimero_derivatives_MI_SF}), a circumstance which can be explained in terms of the delocalizing effect of hopping processes, which tend to smooth down transitions and sharp features of the phase diagram \cite{NoiEntropy,NoiSREP,Noi_Zenesini,NoiPRA3}.

The presence of this staircase-like structure in the central region of the $(\alpha,\,\beta)$ plane is due to the fact that, being the hopping amplitude small, the system responds to small variations of control parameters in a highly non-linear way. As it will be explained in Section \ref{sub:analytic_treatment} by means of a simple analytic treatment, when tunnelling terms tend to zero, phase PL [which, in the CV picture, can be thought of as a collection of states which transform in a \textit{smooth} way when $\alpha$ and $\beta$ are varied] gives way to a sequence of stripes in the $(\alpha,\,\beta)$ plane within which the ground-state configuration proves to be rather rigid upon small variations of $\alpha$ and $\beta$ themselves. The transition between any two such stripes represents an abrupt change in the ground-state configuration and corresponds to the kind of bosons rearrangement pictorially illustrated in Figure \ref{fig:Istogrammi_con_bolle}.

\begin{figure}[h]
\centering
\includegraphics[width=1\linewidth]{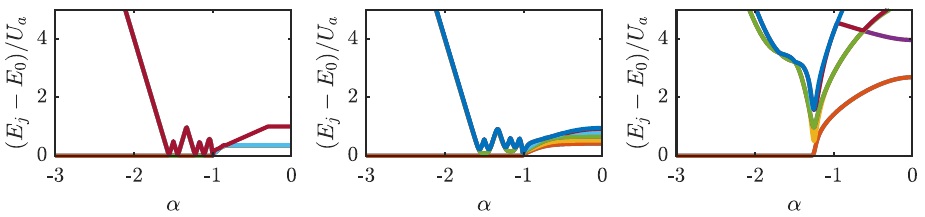}
\caption{First 8 excited energy levels, obtained by means of an exact numerical diagonalization of Hamiltonian (\ref{eq:Hamiltoniana_BH_trimero}), for $T:=T_a=T_b=0,\,0.02,\,0.50$ in the left, central and right panel, respectively. Model parameters $N_a=N_b=15$, $U_a=1$, $U_b=0.36 \,\Rightarrow\, \beta=0.6 $ and $W\in[-1.8,0]\, \Rightarrow \, \alpha \in [-3,0]$ have been chosen.  }
\label{fig:Spectrum_trimero_MI_SF}
\end{figure}

The staircase-like structure corresponding to jerky transfers of bosons from/to the site hosting the supermixed soliton is evident also in terms of the energy fingerprint of the system. The latter, i.e. the set of the first excited energy levels, are shown in Figure \ref{fig:Spectrum_trimero_MI_SF} for different values of the hopping amplitudes. 
In particular, if the hopping amplitudes are sufficiently small (see left and central panel of Figure \ref{fig:Spectrum_trimero_MI_SF}), the energy level structure in the region of the $(\alpha,\,\beta)$ plane between phase $M$ and phase $SM$ features sharp \textit{peaks}. With reference to the aforementioned Figure, where $\beta$ has been set to $0.6$, the staircase-like structure is present for $-1.6\leq \alpha \leq -1$. The number of peaks in the energy spectrum corresponds to that of the stripes that one crosses while walking along a straight line at $\beta=\mathrm{const}$ in the $(\alpha,\,\beta)$ plane. Similarly, the number of valleys visible in the energy spectrum corresponds to that of stripes borders crossed by the constant-$\beta$ pathway. The sequence of stripes whose borders correspond to jerky boson transfers (of the type sketched in Figure \ref{fig:Istogrammi_con_bolle}) can be clearly appreciated also in Figure \ref{fig:Lobi_con_disequazioni}, which has been derived within a fully analytic framework (see Section \ref{sub:analytic_treatment} for details).

If hopping amplitudes $T_a$ and $T_b$ exceed a certain threshold, the discrete character of the interwell boson exchange fades away and the energy levels $E_j(\alpha)$, regarded as functions of control parameter $\alpha$, get well-behaved (see right panel of Figure \ref{fig:Spectrum_trimero_MI_SF}).

Another effective indicator that can provide insight into the jerky transfers of bosons from/to the supermixed soliton is represented by $\mathcal{D}(E_0)$, the degeneracy of the ground-state level when $T_a=T_b=0$. We recall, in this regard, that, as soon as the tunnelling is non-vanishing, the ground state of Hamiltonian (\ref{eq:Hamiltoniana_BH_trimero}) gets unique and not-degenerate \cite{Penna_Wang}, although it can take the form of a superposition of few macroscopically different configurations (a Schr\"{o}dinger-cat state) \cite{Wimberger_Cat,NoiSREP}. As we shall discuss, such a superposition of different Fock states, although being not-degenerate, bears memory of the value of $\mathcal{D}(E_0)$ that one would have if hopping processes were suppressed, since $\mathcal{D}(E_0)$ at $T_a=T_b=0$ corresponds to the number of macroscopic configurations which constitute the non-degenerate Schr\"{o}dinger-cat state at small but finite tunnellings.

\begin{figure}[h]
\centering
\includegraphics[width=0.5\linewidth]{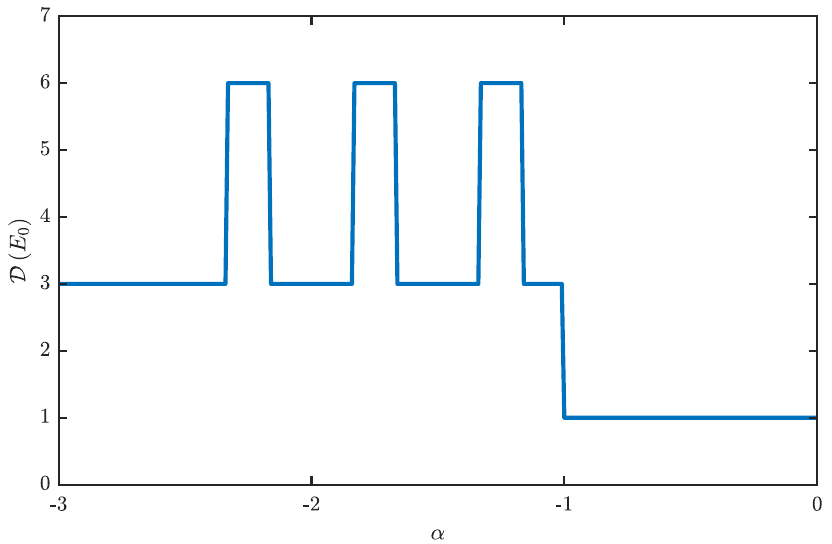}
\caption{ Degeneracy of the ground-state level $E_0$, obtained by means of an exact numerical diagonalization of Hamiltonian (\ref{eq:Hamiltoniana_BH_trimero}), for $T:=T_a=T_b=0$. Model parameters $N_a=N_b=15$, $U_a=1$, $U_b=0.16 \,\Rightarrow\, \beta=0.4 $ and $W\in[-1.2,0]\, \Rightarrow \, \alpha \in [-3,0]$ have been chosen. Each jump discontinuity corresponds to a change in the ground-state structure of the type illustrated in \ref{fig:Istogrammi_con_bolle}.}
\label{fig:Degeneracy}
\end{figure}

The value of $\mathcal{D}(E_0)$, computed along a path in the $(\alpha,\,\beta)$ plane featuring $\beta=\mathrm{const}$, is illustrated in Figure \ref{fig:Degeneracy}. At the chosen value of $\beta$, for $\alpha < 2.3$, the system's ground state takes the form of a supermixed soliton whose degeneracy $\mathcal{D}$ is $3$, because \textit{three} is the number of its possible positions in the trimer. For $-1<\alpha<0$, the configuration which minimizes the (expectation value of) Hamiltonian (\ref{eq:Hamiltoniana_BH_trimero}) is the uniform and mixed one. The latter is such that there are $N_a/3$ species-a and $N_b/3$ species-b bosons in each site. If, as in the case of Figure \ref{fig:Degeneracy}, $N_a$ and $N_b$ are integer multiples of the number of lattice sites, there exists just one state which minimizes energy (\ref{eq:Hamiltoniana_BH_trimero}) and, accordingly, the associated degeneracy $\mathcal{D}(E_0)$ is unitary.       

For $-2.3<\alpha<-1$, the system ground state transforms from the mixed to the supermixed one in such a way that bosons are transferred to the emerging supermixed soliton in the jerky fashion sketched in Figure \ref{fig:Istogrammi_con_bolle}. Accordingly, the degeneracy of the ground-state level alternatively takes the values $3$ and $6$, depending on the number of species-a bosons which are not part of the supermixed soliton. To better clarify this property, we observe that, in the region of the phase diagram corresponding to phase PL, at $T_a=T_b=0$, the ground state of Hamiltonian (\ref{eq:Hamiltoniana_BH_trimero}) is made up of Fock states of the type

\begin{equation}
\label{eq:Emerging_Supermixed_soliton}
    |N_1,\,N_2,\,N_3,\,M_1,\,M_2,\,M_3\rangle = |N_a-N_2-N_3,\,N_2,\,N_3,\,N_b,\,0,\,0\rangle
\end{equation}
where $N_2<N_a-N_2-N_3$ and $N_3<N_a-N_2-N_3$. In the light of this, one can immediately conclude that the degeneracy of the associated energy level, i.e. the number of possible permutations of the quantum numbers that come into play, is $3$ when $N_2=N_3$ and it is $6$ when $N_2\neq N_3$. With reference to Figure \ref{fig:Lobi_con_disequazioni}, which is obtained by means of the fully analytic framework discussed in Section \ref{sub:analytic_treatment}, purple (yellow) stripes are associated to $\mathcal{D}=6$ ($\mathcal{D}=3$), while, as already explained, in regions SM and M, $\mathcal{D}$ takes the values $3$ and $1$, respectively. The fact that purple and yellow stripes have different widths will be explained by the simple analytic framework presented in Section \ref{sub:analytic_treatment}.    

\begin{figure}[h]
\centering
\includegraphics[width=1\linewidth]{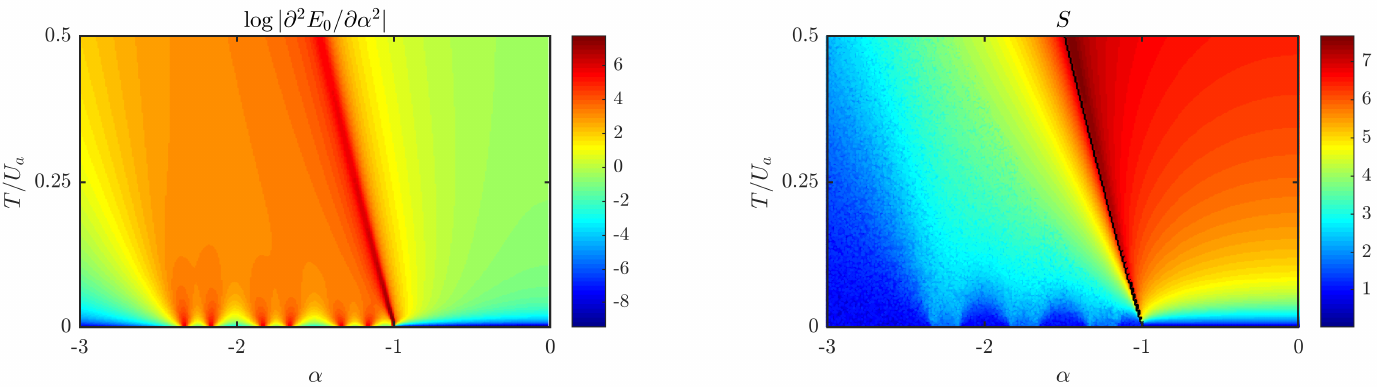}
\caption{The mechanism of jerky interwell boson transfer is present provided that tunnelling $T$ is small enough. Left panel: second derivative of the ground-state energy (\ref{eq:Ground_state_energy}) with respect to control parameter $\alpha$. Right panel: entropy (\ref{eq:S}) of the probability distribution associated to coefficients $|c(\vec{N},\,\vec{M})|^2$ [see formula (\ref{eq:c_k})]. The black line corresponds to the border of the stability region (\ref{eq:Inequality_trimero}) of the mixed configuration. Notice that, unlike Figures \ref{fig:Quantum_indicators_trimero_MI_SF} and \ref{fig:Quantum_indicators_trimero_derivatives_MI_SF}, these plots are referred to the $(\alpha,\,T/U_a)$ plane, instead of the $(\alpha,\,\beta)$ plane.  Model parameters $N_a=N_b=15$, $U_a=1$, $U_b=0.16\,\Rightarrow\, \beta=0.4$, $T_a=T_b=:T \in[0,0.5]$, and $\alpha\in[-3,0]$ have been used. Each plot includes more than 75k points \cite{Hactar}, corresponding to as many numerical diagonalizations of Hamiltonian (\ref{eq:Hamiltoniana_BH_trimero}). }
\label{fig:Sweep_T}
\end{figure}

The discussed mechanism of jerky interwell boson transfer is present not only at $T=0$, but it persists also for finite values of tunnellings. To better illustrate this circumstance, we refer to Figure \ref{fig:Sweep_T}, where we plot the second derivative of the ground state energy (\ref{eq:Ground_state_energy}) with respect to control parameter $\alpha$ (left panel) and the entropy

\begin{equation}
    \label{eq:S}
    S=-\sum_{\vec{N}}\sum_{\vec{M}}|c(\vec{N},\,\vec{M})|^2 \log |c(\vec{N},\,\vec{M})|^2
\end{equation}
of the probability distribution associated to coefficients (\ref{eq:c_k}) (right panel). Both plots, which are referred to the $(\alpha,\,T/U_a)$ plane, clearly show the presence of \textit{lobes} for small values of $T/U_a$ and for $-2.4<\alpha<-1$. More specifically, as regards the left plot, one can appreciate a sequence of \textit{six} lobes (depicted in green) which correspond to a sequence of Fock states of the type

$$
|\psi_0 \rangle_6 \approx \frac{1}{\sqrt{6}}\left[|N_a-N_2-N_3,\,N_2,\,N_3,\,N_b,\,0,\,0\rangle   +   |N_a-N_2-N_3,\,N_3,\,N_2,\,N_b,\,0,\,0\rangle\right. +
$$
$$
  + |N_2,\,N_a-N_2-N_3,\,N_3,\,0,\,N_b,\,0\rangle +     |N_3,\,N_a-N_2-N_3,\,N_2,\,0,\,N_b,\,0\rangle+
$$

\begin{equation}
    \label{eq:Gatto_6}
\left.    + |N_2,\,N_3,\,N_a-N_2-N_3,\,0,\,0,\,N_b\rangle + |N_3,\,N_2,\,N_a-N_2-N_3,\,0,\,0,\,N_b\rangle\right]
\end{equation}
for $N_2 \neq N_3$, and of the type 

$$
|\psi_0 \rangle_3 \approx \frac{1}{\sqrt{3}}\left[|N_a-N_2-N_3,\,N_2,\,N_3,\,N_b,\,0,\,0\rangle + |N_2,\,N_a-N_2-N_3,\,N_3,\,0\,N_b,\,0\rangle + \right.
$$

\begin{equation}
    \label{eq:Gatto_3}
\left.    + |N_2,\,N_3,\,N_a-N_2-N_3,\,0,\,0,\,N_b\rangle\right] +
\end{equation}
for $N_2 = N_3$, where the symbol ``$\approx$" has been used to recall that, when $T>0$, many other Fock states $|\vec{N},\,\vec{M}\rangle$ enter into the expression of $|\psi_0\rangle$, but their weights $|c(\vec{N},\vec{M})|^2$ [see (\ref{eq:c_k})] in the linear combination $|\psi_0\rangle=\sum_{\vec{N}}\sum_{\vec{M}}c(\vec{N},\,\vec{M})|\vec{N},\,\vec{M}\rangle$ is very small if ratio $T/U_a$ is, in turn, small. Going from left to right in both plots of Figure \ref{fig:Sweep_T}, for small enough values of $T/U_a$, the quantum number $N_a-N_2-N_3$, which correspond to the number of species-a bosons in the supermixed soliton, takes the value $15$ for $\alpha<-2.4$ (SM configuration), and the value $5$ for $\alpha >-1$. More interestingly, for $-2.4<\alpha<-1$, it takes the sequence of values 14, 13, 12, 11, 10, 9. Accordingly, the system ground state alternately takes the form of state (\ref{eq:Gatto_6}) and state (\ref{eq:Gatto_3}). This sequence of 6 different ground states corresponds to that of the 6 green lobe-like domains in the bottom part of the left panel of Figure \ref{fig:Sweep_T} and to that of the blue lobe-like domains in the bottom part of the right panel of Figure \ref{fig:Sweep_T}. Notice, in this regard, that the domains corresponding to the cases $N_2=N_3$ are bigger, i.e. they are wider and they persist for bigger values of $T/U_a$. Conversely, the lobes corresponding to the cases $N_2\neq N_3$ are narrower and are more easily disrupted by tunnelling. The different \textit{width} of the lobes for $N_2=N_3$ and of those for $N_2\neq N_3$ will be explained in Section \ref{sub:analytic_treatment} (by means of a simple analytical model), while their different \textit{height} can be explained  by means of an analogy with the superfluid-Mott insulator transition. Note that, also, these two kinds of lobes visible in the right panel of Figure \ref{fig:Sweep_T} alternately take the values $S\approx \log 6$ and $S\approx \log 3$, in that the number of macroscopic components present in the non-degenerate Schr\"{o}dinger-cat-like states of the type (\ref{eq:Gatto_6}) and (\ref{eq:Gatto_3}) bears memory of the degeneracy $\mathcal{D}(E_0)$ of the ground state if the tunnelling $T$ was suppressed.

In order to highlight the analogy with the superfluid-Mott insulator transition, we start by looking at the trimer system as if it was made up of two parts. One corresponds to the site where the supermixed soliton is emerging: it includes $N_a-N_2-N_3$ species-a bosons and $N_b$ species-b bosons. The other part corresponds to the remaining two sites, hosting, in total, $N_2+N_3$ species-a bosons and $0$ species-b bosons. As $N_a-N_2-N_3$ can be much bigger than $N_2+N_3$, the macroscopically occupied site can be thought of as a \textit{reservoir of species-a bosons} and the remaining two sites can be regarded as a two-well system including just one bosonic species (instead of a binary mixture) which is in contact with a particle reservoir. In this perspective, the interspecies attraction $W$, and hence effective control parameter $\alpha$ [see formula (\ref{eq:alfa_beta})], plays the role of an effective \textit{chemical potential}, as it can control the release/absorption of species-a bosons from/to the particle reservoir.

Notice that there is a profound difference between states (\ref{eq:Gatto_3}) and states (\ref{eq:Gatto_6}). Concerning the effective two-well potential resulting from the exclusion of the macroscopically occupied site (which plays the role of particle reservoir), the former are marked by a \textit{commensurate} filling, while the second feature an \textit{incommensurate} filling. As a consequence, lobes corresponding to the case of $N_2=N_3$ play the role of Mott lobes, while those corresponding to the case $N_2\neq N_3$ correspond to \textit{superfuid} lobes, as one species-a boson is shared between the sites of the effective two-well potential.

\begin{figure}[h]
\centering
\includegraphics[width=0.5\linewidth]{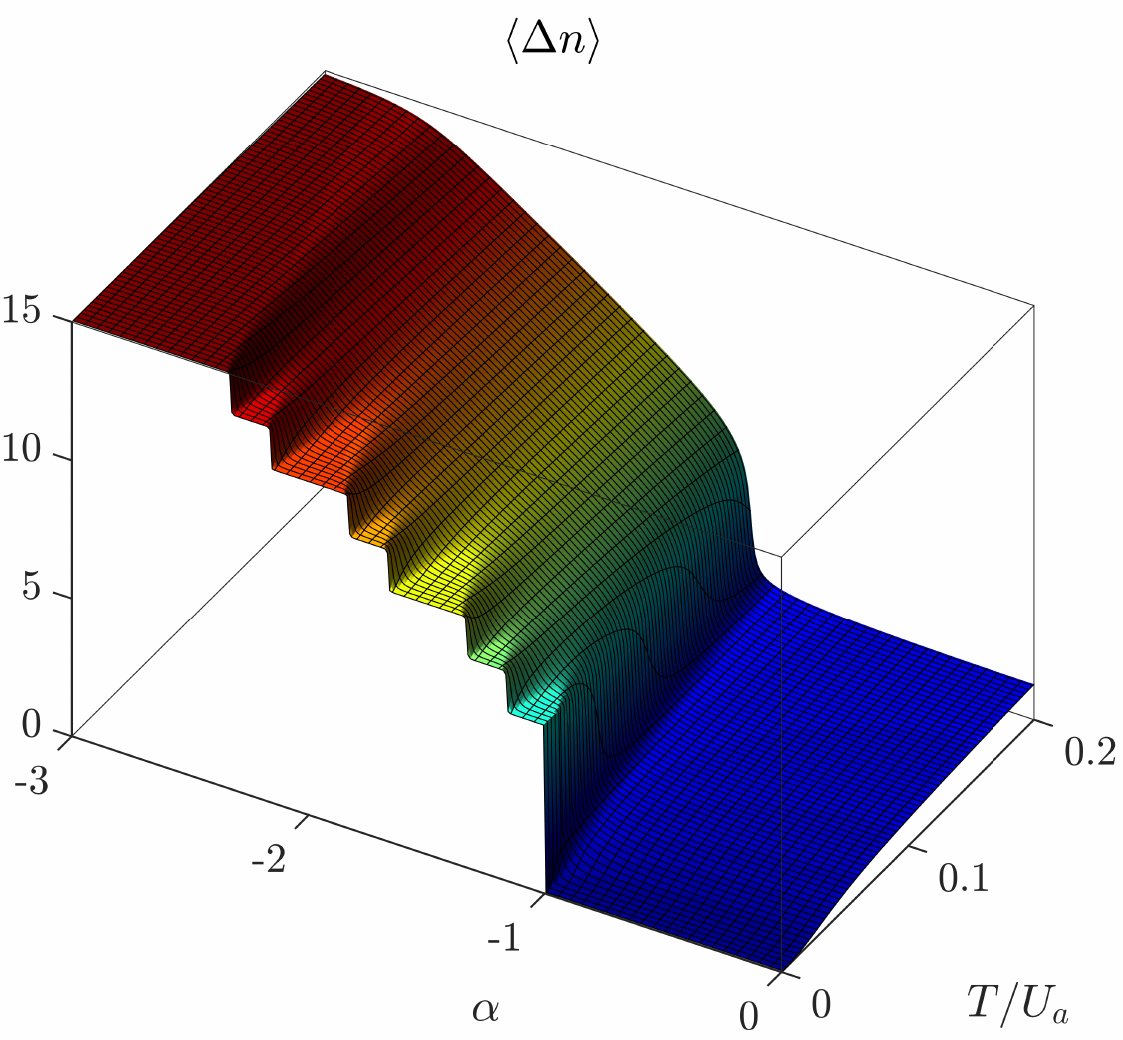}
\caption{Expectation value $\langle \Delta n\rangle =\langle \psi_0|\Delta n |\psi_0\rangle$ of operator imbalance operator $\Delta n$ [see formula (\ref{eq:Delta_n})] as a function of $\alpha$ and $T/U_a$. The mechanism of jerky interwell boson transfer is present provided that tunnelling $T$ is small enough (compare the staircase-like structure for $T/U_a\to 0$ with the slide-like appearance for $T/U_a\approx 0.2$). Notice that, unlike Figures \ref{fig:Quantum_indicators_trimero_MI_SF} and \ref{fig:Quantum_indicators_trimero_derivatives_MI_SF}, these plots are referred to the $(\alpha,\,T/U_a)$ plane, instead of the $(\alpha,\,\beta)$ plane. Notice also that, unlike Figure \ref{fig:Sweep_T}, the range of $T/U_a$ is $[0,\,0.2]$ in order to better appreciate the presence of lobe-like regions. Model parameters $N_a=N_b=15$, $U_a=1$, $U_b=0.16\,\Rightarrow\, \beta=0.4$, $T_a=T_b=:T \in[0,0.5]$, and $\alpha\in[-3,0]$ have been used. The plot includes more than 60k points \cite{Hactar}, corresponding to as many numerical diagonalizations of Hamiltonian (\ref{eq:Hamiltoniana_BH_trimero}). }
\label{fig:Sweep_T_delta_n}
\end{figure}

Interestingly, both states (\ref{eq:Gatto_3}) and states (\ref{eq:Gatto_6}) seem to undergo a deep change when $T$ exceeds a certain threshold, which is different in the two cases ($T/U_a\approx 0.02$ and $\approx 0.01$, respectively). In the first case, as $N_2=N_3$ (commensurate filling), the analogy with the superfluid-Mott insulator transition suggests that, increasing the ratio $T/U_a$, bosons tend to \textit{delocalize} and the system switches from the Mott to the superfluid phase. Concerning the other family of states, (\ref{eq:Gatto_6}), featuring $N_2\neq N_3$, the interpretation is more delicate. This because they are \textit{already} endowed with a superfluid character, as one boson is shared by the two sites of the resulting effective system. Although this property deserves further investigation (we expect that an increasingly rich structure of ``superfluid lobes" necessarily emerges when the number of lattice sites increases), it is possible to conjecture that, crossing the border of such a lobe, the system switches from a weaker to a stronger type of superfluidity. In fact, states (\ref{eq:Gatto_6}) are forcefully superfluid, even for $T\to 0^+$, because of the extra boson expelled by the supermixed soliton and injected into the effective two-well potential. Nevertheless, the superfluid character of state (\ref{eq:Gatto_6}) is strongly dammed by the fact that it includes just 6 Fock states (actually 2, if one neglects the possible ways to permute the position of the particle reservoir) and therefore it is far from being of the type

\begin{equation}
    |\psi_0\rangle \propto (A_2^\dagger+A_3^\dagger)^{N_2+N_3} |0,0\rangle,
\end{equation}
the latter representing the exact ground state of a two-well BH Hamiltonian featuring $U/T\to 0$ and hosting $N_2+N_3$ species-a bosons. This circumstance would reasonably explain the presence of small lobes in both panels of Figure \ref{fig:Sweep_T}. In the same spirit of Ref. \cite{Minguzzi_Hekking}, where suitable squeezing indicators were introduced to detect lobe-like structures in an asymmetric BH-dimer Hamiltonian, we introduce indicator

\begin{equation}
    \label{eq:Delta_n}
    \Delta n = \frac{1}{2}\left(2N_{max}-N_i-N_j\right),
\end{equation}
where $N_{max}:=\max_{k\in\{1,2,3\}}\{N_k\}$ and $N_i,\,N_j \in\{N_1,\,N_2,\,N_3\}-\{N_{max}\}$, which corresponds to the average species-a bosons imbalance between the site hosting the supermixed soliton and the sites of the remaining two-well system. As it is visible in Figure \ref{fig:Sweep_T_delta_n}, where the expectation value $\langle \Delta n\rangle =\langle \psi_0|\Delta n |\psi_0\rangle$ is plotted,  when $T/U_a$ is small enough, a sequence of lobe-like domains is present, which corresponds to the sequence of values 15 (SM configuration), 13.5 (first superfluid-like lobe), 12 (first Mott-like lobe), 10.5, and so on. 

We conclude this Section by recalling that it is possible to find, either within the CV picture \cite{NoiPRA3}, or by means of the dynamical-algebra method \cite{NoiPRA2}, the region of the parameters space where the mixed configuration (the one sketched in the leftmost panel of Figure \ref{fig:Istogrammi}) is stable. It is given by inequality 

\begin{equation}
        \label{eq:Inequality_trimero}
        \alpha > -\sqrt{\left(1+\frac{9}{2}\frac{T_a}{U_aN_a}\right)\left(1+\frac{9}{2}\frac{T_b}{U_bN_b}\right)}
\end{equation}
whose border, in the $(\alpha,T/U_a)$ plane, corresponds to the black line in the right panel of Figure \ref{fig:Sweep_T}. Interestingly, one can notice that, while approaching this border from the right, entropy (\ref{eq:S}) associated to the ground state significantly increases and takes the maximum value exactly at the value of $\alpha$ where the mixed configuration gives way to a configuration of the type (\ref{eq:Emerging_Supermixed_soliton}). 

\subsection{Analytic treatment}
\label{sub:analytic_treatment}
\begin{figure}[h!]
\centering
\includegraphics[width=0.5\linewidth]{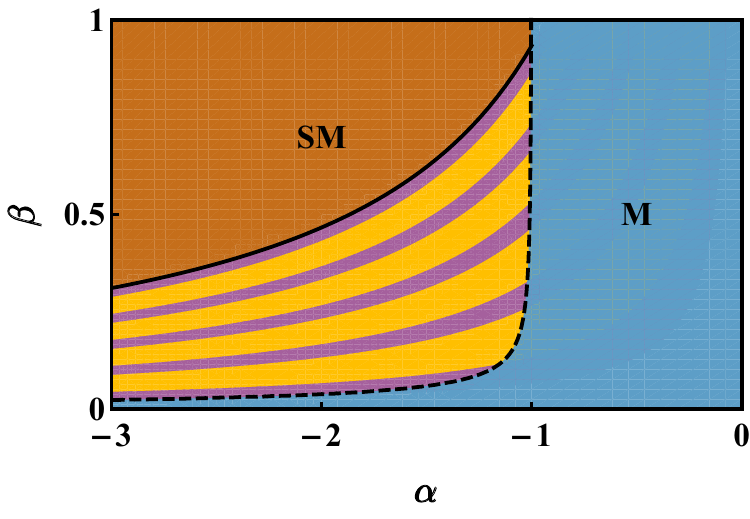}
\caption{Map of the system's minimum-energy configurations. It corresponds to the graphical representation of the set of inequalities derived in Section \ref{sub:analytic_treatment}. More specifically: the solid black [dashed] line corresponds to condition (\ref{eq:Inequality_SM_SM_m1}) [(\ref{eq:alpha_max})], while the set of purple [yellow] stripes is given by condition (\ref{eq:Inequality_odd}) [(\ref{eq:Inequality_even})]. Model parameters $N_a=N_b=15$, $U_a=1$, $U_b\in[0,1]\,\Rightarrow\, \beta\in[0,1]$, $\alpha\in[-3,0]$ and $T_a=T_b=0$ have been used.}
\label{fig:Lobi_con_disequazioni}
\end{figure}

We present a simple but effective analytical treatment, capable of capturing the presence of the staircase-like structure in the central region of the $(\alpha,\beta)$ plane (see Figure \ref{fig:Quantum_indicators_trimero_derivatives_MI_SF}). By means of fully-analytic computations, we derive, for $T=0$, a set of inequalities giving the stability region not only of the mixed and of the supermixed configurations, but also of each intermediate configuration of the type (\ref{eq:Emerging_Supermixed_soliton}). The graphical representation of these inequalities is shown in Figure \ref{fig:Lobi_con_disequazioni}, which effectively mimics the scenario illustrated in Figure \ref{fig:Quantum_indicators_trimero_derivatives_MI_SF}, obtained, in turn, by sweeping model parameters and numerically diagonalizing Hamiltonian (\ref{eq:Hamiltoniana_BH_trimero}).

% Ricollegarsi ai riferimenti a questa sezione fatti nelle sezioni precedenti.

Let us consider a supermixed-soliton configuration. The associated energy, for $T=0$, reads

\begin{equation}
\label{eq:E_SM}
 E(SM)=\frac{U_a}{2}N_a(N_a-1)+\frac{U_b}{2}N_b(N_b-1)+WN_aN_b.
\end{equation}
The first state belonging to the  staircase-like structure differs from a supermixed-soliton state because one species-a boson has left the macroscopically occupied site and has moved to the remaining two-well system. The energy of this configuration reads

\begin{equation}
    \label{eq:E_SM_m1}
    E(SM-1_a)=\frac{U_a}{2}(N_a-1)(N_a-2)+\frac{U_b}{2}N_b(N_b-1)+W(N_a-1)N_b.
\end{equation}
By solving inequality $E(SM)<E(SM-1_a)$ one obtains that the supermixed configuration ceases to be the energetically favorable one for 

\begin{equation}
    \label{eq:Inequality_SM_SM_m1}
     \alpha>\frac{1}{\beta}\left(\frac{1}{N_a}-1\right). 
\end{equation}
This condition corresponds to the solid black line in Figure \ref{fig:Lobi_con_disequazioni} and allows one to recognize the border between the region of SM states and the first element of the staircase-like structure. It is worth mentioning the fact that it would be energetically \textit{unfavourable} to remove a species-b (instead of a species-a) boson from the supermixed soliton. The condition $E(SM-1_a)<E(SM-1_b)$ is indeed always verified in the chosen range $\beta \in [0,1]$ because of the asymmetric role of species-a and species-b parameters in the definition of $\beta$ [see formula (\ref{eq:alfa_beta})]. State $|N_a-1,\,1,\,0,\,N_b,\,0,\,0\rangle$ is the actual system ground state provided that condition (\ref{eq:Inequality_SM_SM_m1}) is satisfied \textit{and} that $E(SM-1_a)<E(SM-2_a)$. The latter inequality corresponds to the border between the upper purple stripe and its neighbouring yellow stripe in Figure \ref{fig:Lobi_con_disequazioni}.

One can easily generalize this reasoning in order to find the condition under which a state of the type (\ref{eq:Emerging_Supermixed_soliton}) and such that $K_a=N_2+N_3$ species-a bosons have left the supermixed soliton, is the actual system's ground state. One needs to distinguish two cases: $K_a$ odd, and $K_a$ even. After some straightforward algebra, it turns out that the aforementioned state, whose energy is $E(SM-K_a)$, is the actual ground state provided that

\begin{equation}
    \label{eq:Inequality_odd}
    \frac{1}{\beta}\left(\frac{3(K_a-1)/2+1}{N_a}-1\right)\quad<\quad \alpha\quad <\quad \frac{1}{\beta}\left(\frac{3(K_a-1)/2+2}{N_a}-1\right) \qquad \text{if $K_a$ is odd},
\end{equation}
\begin{equation}
    \label{eq:Inequality_even}
    \frac{1}{\beta}\left(\frac{3K_a/2-1}{N_a}-1\right)\quad<\quad \alpha\quad <\quad \frac{1}{\beta}\left(\frac{3K_a/2+1}{N_a}-1\right) \qquad \text{if $K_a$ is even}.
\end{equation}
With reference to Figure \ref{fig:Lobi_con_disequazioni}, the former (latter) set of inequalities corresponds to the set of purple (yellow) stripes. Notice also that these simple analytical expressions perfectly capture the fact that yellow stripes are two times wider than purple stripes or, in other words, that, in Figure \ref{fig:Degeneracy}, the pulses with degeneracy $\mathcal{D}(E_0)=6$ are two times narrower than those with degeneracy $\mathcal{D}(E_0)=3$. The same reasoning, of course, accounts for the different widths of superfluid-like and Mott-insulator-like lobes of Figure \ref{fig:Sweep_T} (see the relevant discussion in Section \ref{sub:Exact_results}).

It is known from the theory developed in Ref. \cite{NoiPRA3} and reviewed in Section \ref{sec:Review_PRA3} that, when $\alpha$ approaches the value $\approx -1$, the system ground state sharply switches to the uniform and mixed (M) configuration, featuring $N_a/3$ species-a and $N_b/3$ species-b bosons in each well. This dramatic change in the structure of the ground state corresponds, in the thermodynamic limit, to the transition M-PL (see Section \ref{sec:Review_PRA3}). To derive the condition under which the mixed configuration, featuring energy

\begin{equation}
    \label{eq:Ene_M}
    E(M)= 3 \frac{U_a}{2} \frac{N_a}{3}\left(\frac{N_a}{3}-1\right)   +3\frac{U_b}{2} \frac{N_b}{3} \left(\frac{N_b}{3}-1\right)+3W \frac{N_a}{3}\frac{N_b}{3},
\end{equation}
gets energetically favourable, one needs to solve the inequality $E(M)<E(SM-K_a)$, giving 

\begin{equation}
    \label{eq:alpha_K}
     \alpha > \left(\frac{3K_a}{4N_a} - \frac{1}{2}\right)\frac{1}{\beta} - \frac{N_a\beta}{2N_a - 3K_a},
\end{equation}
and then impose that the critical value of $\alpha$ falls exactly where the lobe with energy $E(SM-K_a)$ would give way to the lobe with energy $E(SM-K_a-1)$. As a result, one obtains relation

\begin{equation}
    \label{eq:alpha_max}
    \alpha^* = -\frac{\sqrt{\beta^2 N_a^2+1}}{\beta N_a}
\end{equation}
giving the border between region M and the staircase-like structure (see black dashed line in Figure \ref{fig:Lobi_con_disequazioni}) and relation

\begin{equation}
    \label{eq:K_a_max}
    K_{a,max}=\frac{2}{3} \left(N_a-1-\sqrt{\beta^2 N_a^2+1}\right)
\end{equation}
giving, for a certain value of $\beta$, the maximum number of species-a bosons which can be subtracted from the supermixed soliton before abruptly switching to the uniform and mixed configuration (of course, as $K_{a,max}$ must be an integer number, the use of the floor function is implicitly needed). 

It is important to remark that the presence of the staircase-like structure which is observed for small values of $T$ and finite boson populations, $N_a$ and $N_b$, is \textit{not} in contrast with the analysis developed within the CV picture (see Ref. \cite{NoiPRA3} and its brief review in Section \ref{sec:Review_PRA3}), but it is complementary to it. In fact, in the limit of large boson populations, one loses track of the quantum granularity which is responsible for the sequence of superfluid-like and Mott-insulator-like lobes and one re-obtains the same expressions which were obtained by approximating boson populations with continuous variables. For example, one has that  

$$
   \lim_{N_a \to +\infty } \alpha^* = -1,
$$
which corresponds to the M-PL border in the phase diagram illustrated in Figure \ref{fig:Diagramma_di_fase}, and also 

$$
\lim_{N_a\to + \infty} \frac{K_{a,max}}{N_a} = \frac{2(1-\beta)}{3},
$$
which perfectly matches the results obtained within the CV picture (see Table \ref{tab:Fasi_trimero} at the M-PL transition).

%%%%%%%%%%%%%%%%%%%%%%%%%%%%%%%%%%%%%%%%%%
\section{Conclusions}
In this work, we have investigated the quantum-granularity effect characterizing the formation of supermixed solitons in ring lattices. It occurs for small values of the tunnelling parameters and consists in a jerky transfer of bosons from/to the site hosting the supermixed soliton. Interestingly, we have shown that it is possible to draw an analogy between the physics of a mixture trapped in a few-well potential and that of the superfluid-Mott insulator transition. More specifically, we have shown that, in certain regimes, the interspecies attraction plays the role of an effective chemical potential and therefore controls the release of bosons from a macroscopically occupied site which, in turn, plays the role of particle reservoir.

In Section \ref{sec:The_model}, we have introduced the model, highlighting the fact the we are considering a bosonic binary mixture featuring repulsive intraspecies and attractive interspecies couplings. In Section \ref{sub:system_phase_diagram}, we have presented the system phase diagram, which was shown to be spanned by just two effective parameters, accounting for the ratio between inter- and intraspecies couplings, and incorporating the possible asymmetry between bosonic species. Section \ref{sub:Quantum_indicators} has been devoted to the presentation of several quantum indicators which are conveniently used to quantify the degree of localization and mixing of the two bosonic species, and the amount of quantum correlation (entanglement) between them. In Section \ref{sec:Beyond} we have pointed out that small hopping amplitudes are responsible for a discrete interwell boson exchange and hence for the emergence of a staircase-like structure in the central region of the phase diagram. To this purpose, in Section \ref{sub:Exact_results}, we have shown the behaviour of different quantum indicators including but not limited to the energy spectrum, various types of entropy, and the degree of degeneracy of the ground-state level. The interesting analogy with the mechanism of the superfluid-Mott insulator transition has been also discussed. Eventually, in Section \ref{sub:analytic_treatment} we have presented a simple but effective analytic framework capable of capturing the emergence of the quantum-granularity effect and the ensuing properties. The rich sequence of Mott-like and superfluid-like lobes revealed for the ring trimer is expected to be present in larger-size lattices. This aspect deserves further investigation which we shall develop elsewhere.

%%%%%%%%%%%%%%%%%%%%%%%%%%%%%%%%%%%%%%%%%%
\paragraph{Acknowledgments.}Both authors warmly thank one the anonymous Referees of Ref. \cite{NoiPRA3} for prompting us to investigate the system in the low-tunnelling regime and in the presence of a finite number of bosons.

%%%%%%%%%%%%%%%%%%%%%%%%%%%%%%%%%%%%%%%%%%
\end{document}